\documentclass[12pt]{article}
\usepackage{mathtext}
\usepackage{graphicx}
\usepackage{booktabs}
\usepackage[english]{babel}
\usepackage[T2A]{fontenc}
\usepackage[utf8]{inputenc}
\usepackage{amsfonts}
\usepackage{amssymb}
\usepackage{amsmath}
\usepackage{pgfplots}
\usepackage{braket}
\usepackage{tikz}
\usepackage{float}
\renewcommand{\d}{\delta}

\usepackage{amsthm} 
\usepackage{arydshln}
\usepackage{hhline}

\numberwithin{equation}{section}
\numberwithin{example}{section}
\numberwithin{question}{section}
\numberwithin{Remark}{section}
\numberwithin{theorem}{section}
\numberwithin{definition}{section}
\numberwithin{proposition}{section}

\begin{document}
	\title{Description of the non-Markovian dynamics of atoms in terms of a pure state}
	
	\author{Yuri Ozhigov (1,2), You Jiangchuan (1)\\
		{\it 1. Lomonosov Moscow State University,}\\{VMK Faculty, Moscow, Russia
		}
		\\
		{\it 2. Valiev Institute of Physics and Technology of Russian}\\{Academy
			of Sciences, Moscow, Russia}
		\\ 
	}
	\maketitle
	
	\begin{abstract}
		The quantum master equation (QME), used to describe the Markov process of interaction between atoms and field, has a number of significant drawbacks. It is extremely memory intensive, and also inapplicable to the case of long-term memory in the environment. An iterative algorithm for modeling the dynamics of an atomic system in the extended Tavis-Cummings model in terms of a pure state is proposed. The correctness of this algorithm is shown on the example of the interaction of an atomic system with the environment through the exchange of photons with the preservation of coherence. This algorithm is applicable to a wide class of processes associated with photonic machinery, in particular, to chemical reactions.
	\end{abstract}
	
	\section{Introduction}
	
	The task of building a computer simulator of chemistry has not yet been solved. It presents a real challenge to computational mathematics due to the exponentially growing complexity of computations. Its difference from the so-called molecular dynamics lies in the key role of the electromagnetic field, which sharply complicates the picture of the reaction. The literature deals with special objects convenient for numerical analysis - polaritons, which are entangled states of molecular structures and an electromagnetic field (see, for example, \cite{Polariton1}). This representation makes it possible to effectively investigate rather subtle effects such as polariton rotation, photon blockade, and the role of the ambient optical cavity within the Jaynes-Cummings finite-dimensional model (see \cite{Polariton2} and \cite{Polariton3}).
	
	The direction of our work is more related to collective effects and complex chemical scenarios than to the detailed study of fine microscopy, as in the cited papers. Polyatomic systems from the quantum point of view were considered, for example, in \cite{Polariton4}, within the framework of the Holstein-Tavis-Cummings HTC model (see \cite{Polariton5}, \cite{Polariton6}, \cite{Polariton7}).
	
	Our goal is to explore even more complex scenarios than purely chemical ones; more related to biology. Therefore, we need to further simplify our model compared to HTC. In this paper, we consider abstract atoms, calling them {\it artificial} in sign of the fact that they retain only the main attributes of real atoms - electrons responsible for covalent bonds and nuclear dynamics within an integer spatial lattice, and molecular structures composed of them.
	Consideration of large ensembles of such atoms and fields (artificial polaritons) requires very economical expenditure of computational resources and appropriate mathematical approaches.
	
	The first method, which will be discussed, is of a general nature and is associated with the standard description of decoherence in the form of a quantum master equation
	
	\begin{equation}
		\label{master_equation}  
		i\hbar\dot{\rho}=[H,\rho]+i{\cal L}(\rho),\ \ {\cal L}(\rho)=\sum\limits_{i=1}^{N-1}\gamma_i(A_i\rho A_i^+-\frac{1}{2}\{ A_i^+A_i\rho,\rho A^+A\}),
	\end{equation}
	onto the density matrix $\rho(t)$ of the system under consideration. Here, the violation of coherence is represented as the influence of the environment, and this influence enters it in the form of decoherence factors $A_i$, which form an orthonormal basis of the space of Liouville operators, and their specific form does not depend in any way on the Hamiltonian $H$ (\cite{BP}).

	\section{Evolution of the state vector instead of the quantum master equation}
	
	The equation \eqref{master_equation} works great for simple processes that can be roughly considered Markovian, but for chemical interactions, especially for complex dynamic scenarios, it is completely unsuitable. The first reason is the high cost of any method for solving the equation \eqref{master_equation}: to obtain a probability distribution, we need to operate on an array that is quadratic in size. The second reason is that the representation of a state in the form of a density matrix involves the division of probability into classical and quantum, which introduces artificial disorder into the description of the system. But even more important is the third reason - non-observance of the uncertainty relation "time - energy".
	
	Let us explain the last argument with a simple example, when all $A_i$ decoherence factors are reduced only to photon leakage from the cavity: $A_i=a$. Let's take two moments of time $t_1$ and $t_2>t_1$ and suppose that the photon has left the cavity at the moment $t_1$. The received part of the state immediately becomes an object of decoherence and stops interference interaction with the rest of the state. So if the photon also flew out of the cavity at time $t_2$, the resulting state will not interfere in any way with the result of the photon's escape at time $t_1$, although in both cases the energy of the resulting states will be the same. However, if we are talking about single photons, as in the models we are considering, the uncertainty relation $\d t\cdot \d E$ means that we must allow such interference, since the frequency of a photon in finite-dimensional QED models is fixed, and the dispersion of its departure time from cavity is large. The results of the events: the departure of a photon at time $t_1$ and at time $t_2$ can interfere and we must take into account this possibility.
	
	Therefore, for chemical interactions, instead of the equation \eqref{master_equation}, it is advisable to use the evolution of the state vector, which is shown below.
	
	\section{Iterative algorithm for photonic machinery}
	
	By photonic machinery we mean the process associated with the exchange of photons between different atoms in an ensemble distributed in space. The Tavis-Cummings-Hubbard model does not give a complete description of such a process, since in this model the photon trajectories are clearly fixed by a system of waveguides between optical cavities; meanwhile, in free space there is no such fixation.
	
	So, we consider the case when the only factor of decoherence is the leakage of a photon from the cavity: $A=a$.
	
	Let us introduce one new register into the basic state -- the number $m$ of photons emitted from the cavity. Thus, the basic state will look like
	\begin{equation}
		\label{new_basic_state}
		|n,m,atoms\rangle
	\end{equation}
	where $n$ is the number of photons located near the atomic system (inside the conditional cavity) that can interact with the considered atoms, $m$ is the number of distant photons that cannot interact with the system of atoms, and the last register is the states of all atoms in the system. Consider first the simplest case of a diatomic system whose total energy is limited by $\hbar\omega$, so that the basis state will be $|n,m,at_1,at_2\rangle$, where $n+m+at_1+at_2=1 $.
	
	The Hamiltonian of such a system ''atoms + photons'' will have the following form
	\begin{equation}
		\label{ham}
		H=\hbar\omega(a^+a+a_{out}^+a_{out}+\bar\sigma^+\cdot\bar\sigma)+g(a\bar\sigma^++a^+\bar\sigma),\ \bar\sigma=\sigma_1+\sigma_2,
	\end{equation}
	where $a^{(+)}$ are field operators inside the cavity, $a^{(+)}_{out}$ are field operators outside the cavity, $\sigma$ are atomic relaxation (excitation) operators, $\cdot $ stands for dot product. The field, therefore, is actually divided into 2 components - near and far, so that only the first one interacts with the atomic ensemble, but the energy of the second, nevertheless, enters the Hamiltonian.
	
	Now the step of the iterative process that determines the evolution in time of our system will consist of two actions applied to the state $|\Psi(t)\rangle$, the first of which consists in applying the unitary evolution operator, and the second expresses the process of ''interaction with environment''. These actions look like this.
	
	1. Let 
	\begin{equation}
		|\tilde\Psi(t+dt)\rangle=(1-\frac{i}{\hbar}H\ dt)|\Psi(t)\rangle=|0,0,\Phi_{00}\rangle+|0,1,\Phi_{01}\rangle+|1,0,\Phi_{10}\rangle,
		\label{step1}
	\end{equation}
	where $|\Phi\rangle$ denote different atomic states. In our case of total energy $\hbar\omega$ we have: $|\Phi_{00}\rangle=\lambda_{01}|01\rangle+\lambda_{10}|10\rangle$, $|\Phi_{01}\rangle=\mu_{01}|00\rangle$, $|\Phi_{01}\rangle=\nu_{01}|00\rangle$, where $\lambda, \mu$ and $\nu$ -- complex numbers.
	
	2. Let's put
	\begin{equation}
		\label{decoh}
		|\Psi(t+dt)\rangle=|0,0,\Phi_{00}\rangle+\alpha|0,1,\Phi_{01}+\Phi_{10}\rangle,\ \alpha=\frac{\sqrt{\|\Phi_{01}\|^2+\|\Phi_{10}\|^2}}{\|\Phi_{01}+\Phi_{10}\|}.
	\end{equation}
	
	Since all three terms in the expansion of $|\tilde\Psi(t+dt)\rangle$ from item 1 are orthogonal, the choice of $\alpha$ in item 2 ensures that the unit norm of the vector $|\Psi(t)\rangle$ is preserved in all points in time. Pair of states: $|1,0,...\rangle$ and $|0,1,...\rangle$ in $|\tilde\Psi(t+dt)\rangle$ and $|\Psi(t +dt)\rangle$, respectively, can be called a donor and an acceptor in the state evolution step, since the amplitude of the first is summed with the amplitude of the second. Moreover, if the donor norm is of order $dt$, then the acceptor amplitude grows, so that the denominator in \eqref{decoh} is non-zero.
	
	This iterative process is naturally generalized to the case of an influx of photons and to the case of an arbitrary total energy of the system, as well as to the case of multilevel atoms, moreover, for ensembles consisting of atoms of different types of spectrum.
	
	Let us have $n$ atoms $at_1,at_2,...,at_n$ so that each $at_i$ has some spectral transition graph $G_i$. Let also $b_1,b_2,...,b_k$ be all possible photon modes corresponding to all transitions of graphs $G_i,\ i=1,2,...,n$. Then the basic state of the field will look like
	\begin{equation}
		\label{field}
		|n_1,n_2,...,n_k; m_1,m_2,...,m_k\rangle
	\end{equation}
	where $n_j,m_j$ are the numbers of photons of the $b_j$ mode, which are near the considered location and far from it, respectively. We impose on these states the condition $n_i+m_j=const_j$ for each $j=1,2,...,k$. We will call a pair of states $|ph_{don}\rangle,\ |ph_{ac}\rangle$ of the form \eqref{field} donor-acceptor if one of these states is obtained from the other under the action of the operator $A=a_{j_1} a_{j_2}^+$, where $j_1,j_2\in\{ 1,2,...,j\}$, which we will call the operator of this pair.
	
	Let us choose for each donor-acceptor pair $P=(|ph_{don}\rangle,\ |ph_{ac}\rangle)$ some probability $p(P)$.
	
	Now the iterative process described above will be modified as follows. Item 1 will remain as before, while item 2 will be transformed as follows. With probability $p=p(P)$ we choose some pair $P$ from donor-acceptor pairs, and let the operator of this pair map $|ph_{don}\rangle$ to $|ph_{ac}\rangle$. We represent the result of item 1 in the form
	\begin{equation}
		\label{modi}
		|\tilde\Psi(t+dt)\rangle=...+|ph_{don},\Phi_{don}\rangle+|ph_{ac},\Phi_{ac}\rangle+...
	\end{equation}
	where the field part of all states denoted by ellipsis is orthogonal to the field part of both states from $P$. Then we put
	\begin{equation}
		\label{res}
		|\Psi(t+dt)\rangle=...+\alpha |ph_{ac},\Phi_{ac}+\Phi_{don}\rangle+...
	\end{equation}
	where 
	$$
	\alpha=\frac{\sqrt{\|\Phi_{don}\|^2+\|\Phi_{ac}\|^2}}{\|\Phi_{don}+\Phi_{ac}\|}.
	$$
	
	Note that, as a rule, the photon occupancy numbers will be only from the set $\{ 0,1\}$, since the first-order unitary evolution approximation from point 1 cannot give more than one "internal" photon. At the same time, the transformation of an "external" photon into an "internal" photon should occur with a low probability: otherwise, the influx of photons from the outside would exceed their outflow, which would mean an unlimited increase in temperature in the vicinity of the atomic system (\cite{Oz3}) .
	
	\section{The case of a distributed atomic system}
	The above algorithm describes the evolution of an open quantum system localized in space; for example, in a low-quality optical cavity, or simply in a small area of empty space. Here we separate the photons into ''inner'' and ''outer'', so that we have the possibility to also return the ''outer'' photons, turning them into ''inner''.
	
	This algorithm can be generalized to the case of several locations, when the entire system of atoms will be divided into groups $At_1,At_2,...,At_d$, so that within one group the atoms will absorb or emit identical ''internal'' photons for this group, but for different groups ''internal'' photons will be different. At the same time, ''outer'' photons will be identical for different $At_j$ groups. This means that the basic state of the entire distributed system will look like
	$$
	|\bar n_1,\bar n_2,...,\bar n_d,\bar m; \bar{At}\rangle
	$$
	where $\bar n_j=(n_1^j,n_2^j,...,n_k^j)$ - occupation numbers of ''internal'' photons corresponding to the location $j$, $\bar m =(m_1,m_2,...,m_k)$ - occupation numbers of ''external'' photons, $\bar{At}_i$ - atomic system in location $i$. 
	
	Now the evolution of the entire distributed system can be represented through an obvious modification of the above algorithm. Let $P_1,P_2,...,P_d$ be the probability distributions on photon modes corresponding to the atomic groups $At_1,At_2,...,At_d$, obtained from the single probability distribution ${\cal P}$. Then point 2 should be modified as follows: its transformation should refer to the atomic location $j$ with the probability distribution $P_j$.
	
	The physical distance, as well as the optical properties of the medium between atomic locations, should be reflected in the form of a change in the probability distributions $P_j$. Thus, our algorithm will be applicable to atomic systems distributed in space.
	
	Our algorithmic scheme is well suited for parallelization on the nodes of the computing system, which can store the common entangled state $|\Psi(t)\rangle$. In this case, each node stores a part of the state related to one atomic location $At_j$, and the shared memory stores a part of the state related to ''external'' photons: $m_1,m_2,...,m_k$. Such a computing system actually imitates the real world, with the only limitation that it is created manually from nanostructures capable of maintaining entangled states. The maximum proximity of such a system to a real prototype makes the quantum model very valuable in practice.
	
	On the other hand, working with a pure state instead of a mixed state allows one to numerically investigate the behavior of a real system in a given model on a completely classical computer.
	
	\section{Experiments and evaluation of iterative algorithms}
	We compared our algorithm and Euler's iterative method for calculating the density matrix in the classical TCH model and the TCH model, in which atoms can move between cavities. The comparison criterion used by us consists in separating the dark component from the $|\Psi(0)\rangle$ atomic state, which must inevitably arise when photons leak from the cavity.
	
	The iterative Euler method(\cite{Quantum computer}) for calculating the density matrix is as follows.
	
	1. The unitary dynamics of the density matrix is calculated:
	\begin{equation}
		\widetilde{\rho}(t+dt) = e^{-iHdt/\hbar} \rho(t) e^{iHdt/\hbar}.
		\label{equation5.1}
	\end{equation}
	2. The action of the Lindblad superoperator $\mathcal{L}$ on the intermediate density matrix $\rho(t+dt)$ is calculated:
	\begin{equation}
		\rho(t+dt) = \widetilde{\rho}(t+dt) + \frac{1}{\hbar} \mathcal{L}(\widetilde{\rho}(t+dt))dt.
		\label{equation5.2}
	\end{equation}
	
	During the experiment, we used the formula $exp(X) = \sum_{k=0}^{\infty} \frac{1}{k!} X^k$, where $X=-\frac{i}{\hbar}H\ dt$, and refined it up to the 4th order to avoid possible errors caused by the equation \eqref{step1}. We first experimented with the photon model mentioned in Section 3 and obtained the following results.
	
	\begin{figure}[H]
		\centering
		\begin{minipage}[t]{0.49\linewidth}	
			\centering
			\includegraphics[width=3.3in]{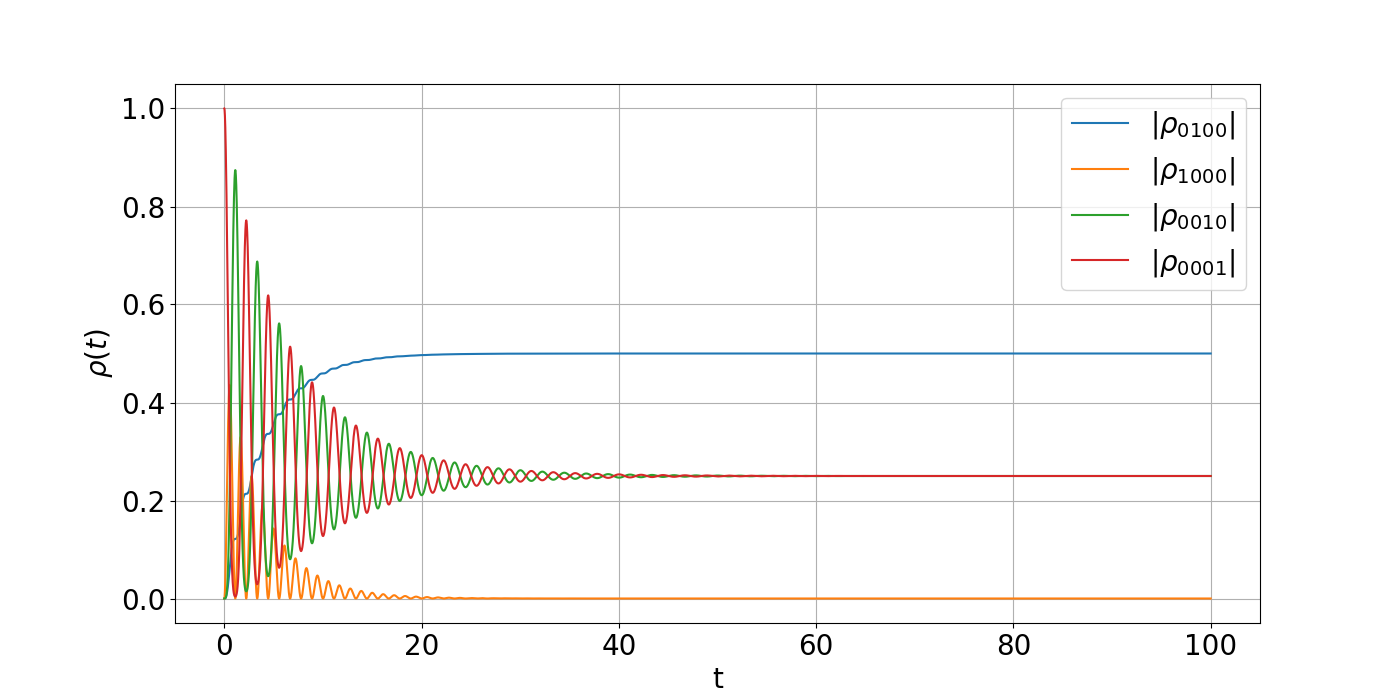}
			Calculation using density matrix. $\gamma=0.5$.
			\label{fig:test_rho_1.png} 
		\end{minipage}\begin{minipage}[t]{0.49\linewidth}
			\centering
			\includegraphics[width=3.3in]{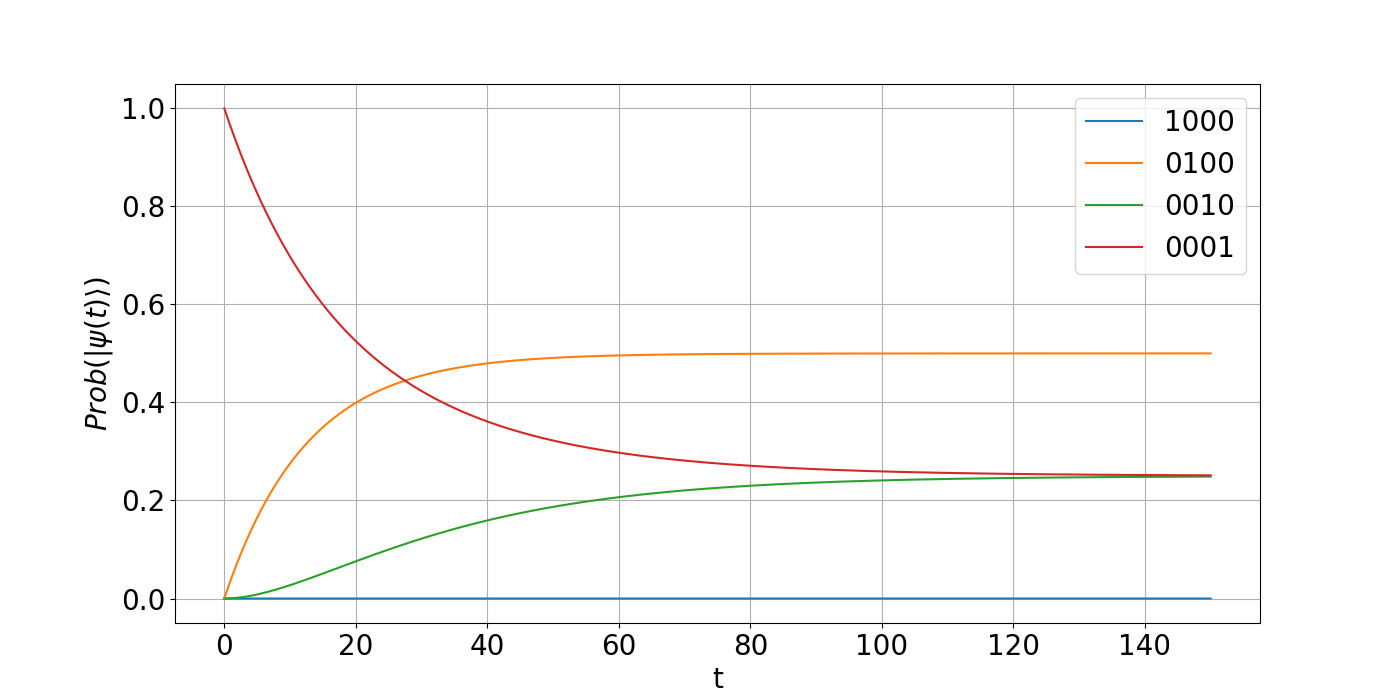}
			Calculation using state vectors.
			\label{fig:test_psi_1_15000_leak.png} 
		\end{minipage}
		\caption{The case of two two-level atoms in a cavity. $dt=0.01, \hbar=1, g=2, \omega = 1$ and initial state $|\Psi(0)\rangle=|00\rangle|01\rangle$. The total number of possible states is 4.}
	\end{figure} 
	Obviously, we got the same result (highlighting the dark state $|D\rangle=\frac{1}{2}|00\rangle$ $\otimes\left(|01\rangle - |10\rangle\right)$ ) using the state vector calculation. Dark states are called states of atomic ensembles, being in which atoms cannot emit light, despite the nonzero energy of excitations. Darkness is an interference effect of a purely quantum nature, which has a significant effect on the interaction of light and matter (see \cite{AL}-\cite{1}). For the Tavis-Cummings model, the dimension of the dark subspace and the explicit form of the dark states are defined in \cite{Oz1},\cite{Oz2}). But the curve converges more slowly than when using the density matrix. In addition, if $dt$ is removed in the second iteration of the Euler \eqref{equation5.2} method and $\gamma=1$ is set, then the resulting image will be consistent with the image obtained using the state vector.
	
	We then increased the maximum number of photons to two: $n+m+atom_1+atom_2 = 2$ to make sure our algorithm also works well for multiphoton systems.
	\begin{figure}[H]
		\centering
		\begin{minipage}[t]{0.49\linewidth}	
			\centering
			\includegraphics[width=3.3in]{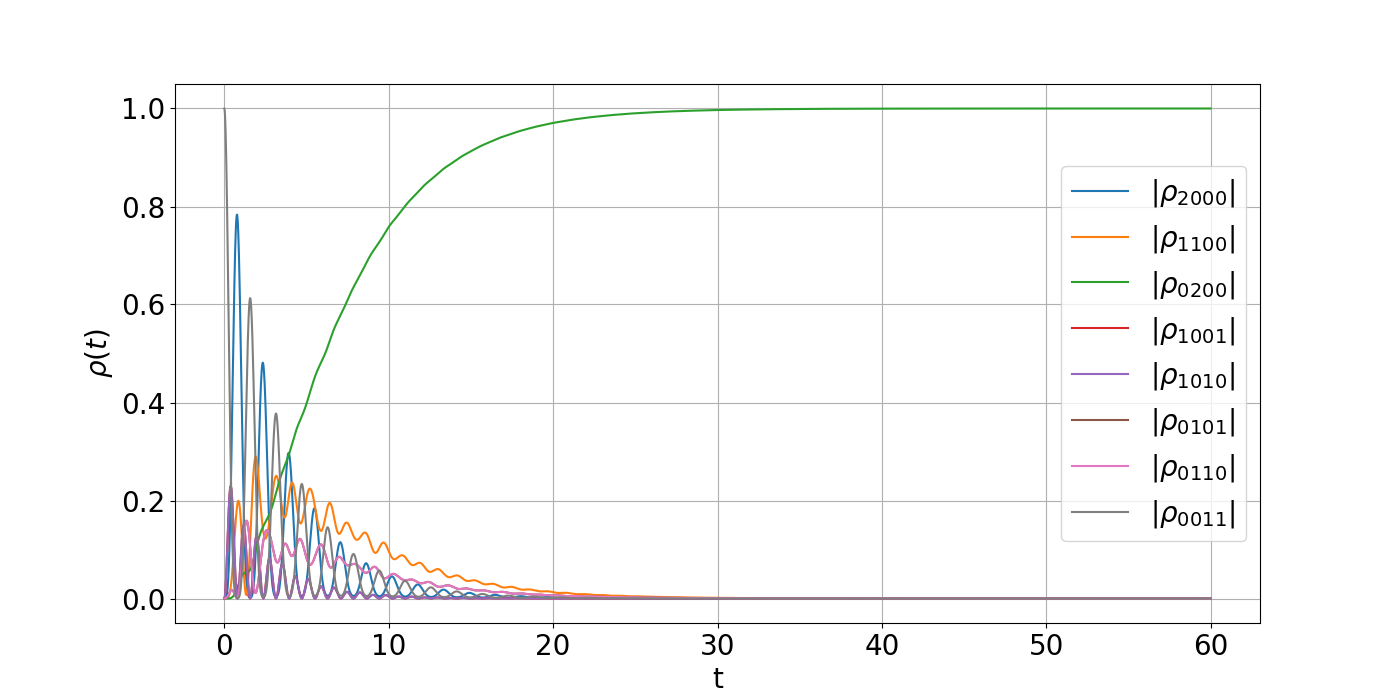}
			Calculation using density matrix. $\gamma=0.5$.
			\label{fig:test_rho_2.png} 
		\end{minipage}\begin{minipage}[t]{0.49\linewidth}
			\centering
			\includegraphics[width=3.3in]{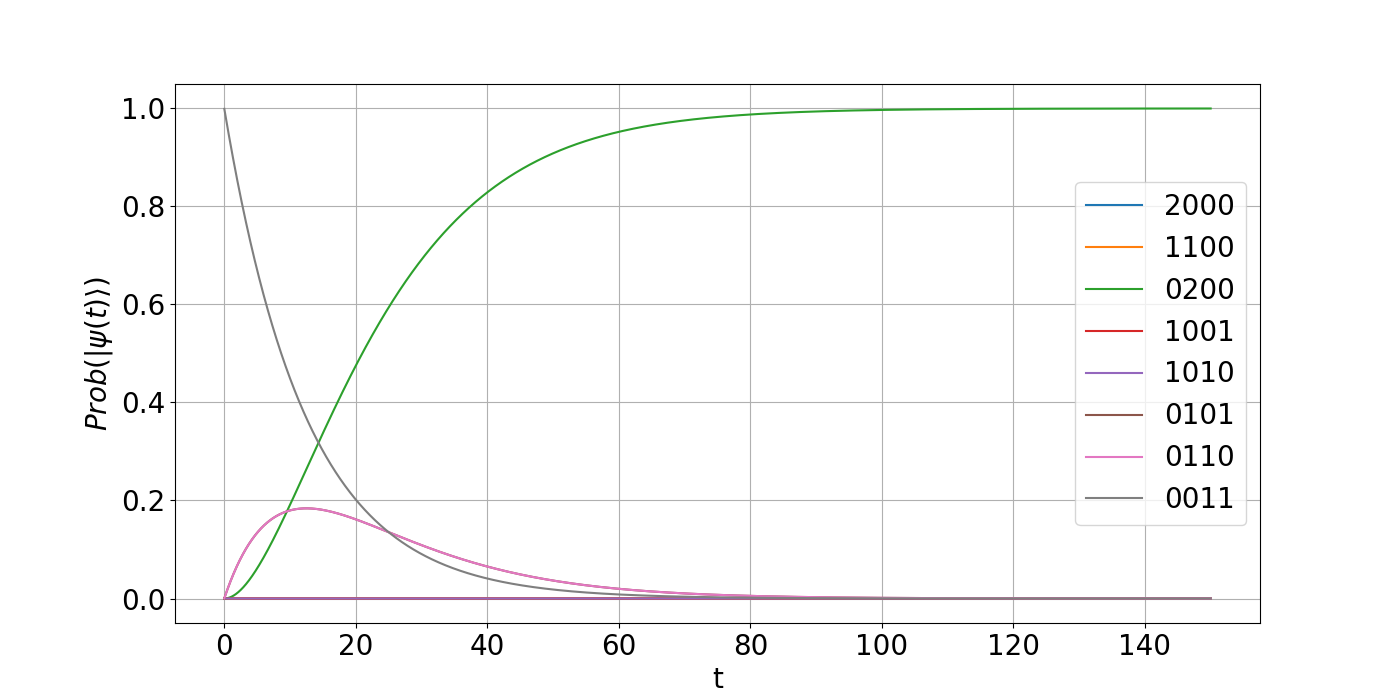}
			Calculation using state vectors.
			\label{fig:test_psi_2_15000_leak.png} 
		\end{minipage}
		\caption{The case of two two-level atoms in a cavity. $dt=0,01, \hbar=1, g=2, \omega = 1$ and initial state $|\Psi(0)\rangle=|00\rangle|11\rangle$. The total number of possible states is 8, there are no dark states.}
	\end{figure} 
	
	We also considered an experiment in which atoms can move freely in a cavity like photons under two cavities, and the number of atoms is constantly increasing. Then the number of possible states will grow exponentially. For example, the base state for two atoms in two connected cavities is:
	\begin{equation}
		\begin{aligned}
			|\Psi\rangle = |n_1n_2\rangle|m_1m_2\rangle|atom_1,atom_2\rangle|position_1,position_2\rangle,\\
			n_1+n_2+m_1+m_2+atom_1+atom_2 = 1.
		\end{aligned}
		\label{equation5.3}
	\end{equation}
	In the general case, the Hamiltonian should be supplemented with a term corresponding to the displacements of atoms between the cavities:
	\begin{equation}
		H_{tun} = \sum_{i,1\le j<q\le k} r_{jq}^i(S(i)_j^+ S(i)_q+S(i)_q^+S(i)_j).\label{formula5.4}
	\end{equation}
	where $r_{jq}^i$ are the non-negative intensities of the tunneling of the atom $i$ from the cavity $j$ to the cavity $q$, $S(i)_j^+, S(i)_j$ are the creation and annihilation operators atom $i$ in cavity $j$:
	\begin{equation}
		S(i)_j^+S(i)_q:|positon_i=q\rangle \to |positon_i=j\rangle
	\end{equation}
	\begin{figure}[H]
		\centering
		\begin{minipage}[t]{0.49\linewidth}	
			\centering
			\includegraphics[width=3.3in]{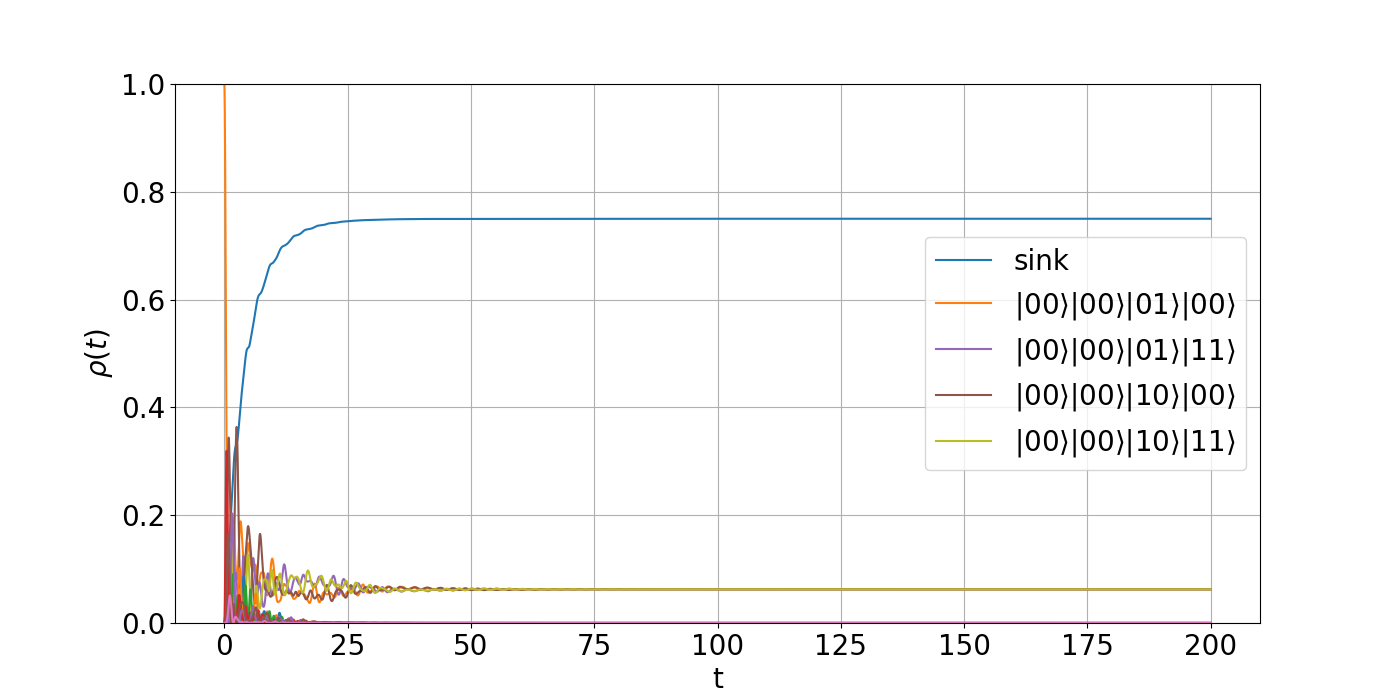}
			Calculation using density matrix. $\gamma=0.5$.
			\label{fig:test_TCH_rho_1_20000leak.png} 
		\end{minipage}\begin{minipage}[t]{0.49\linewidth}
			\centering
			\includegraphics[width=3.3in]{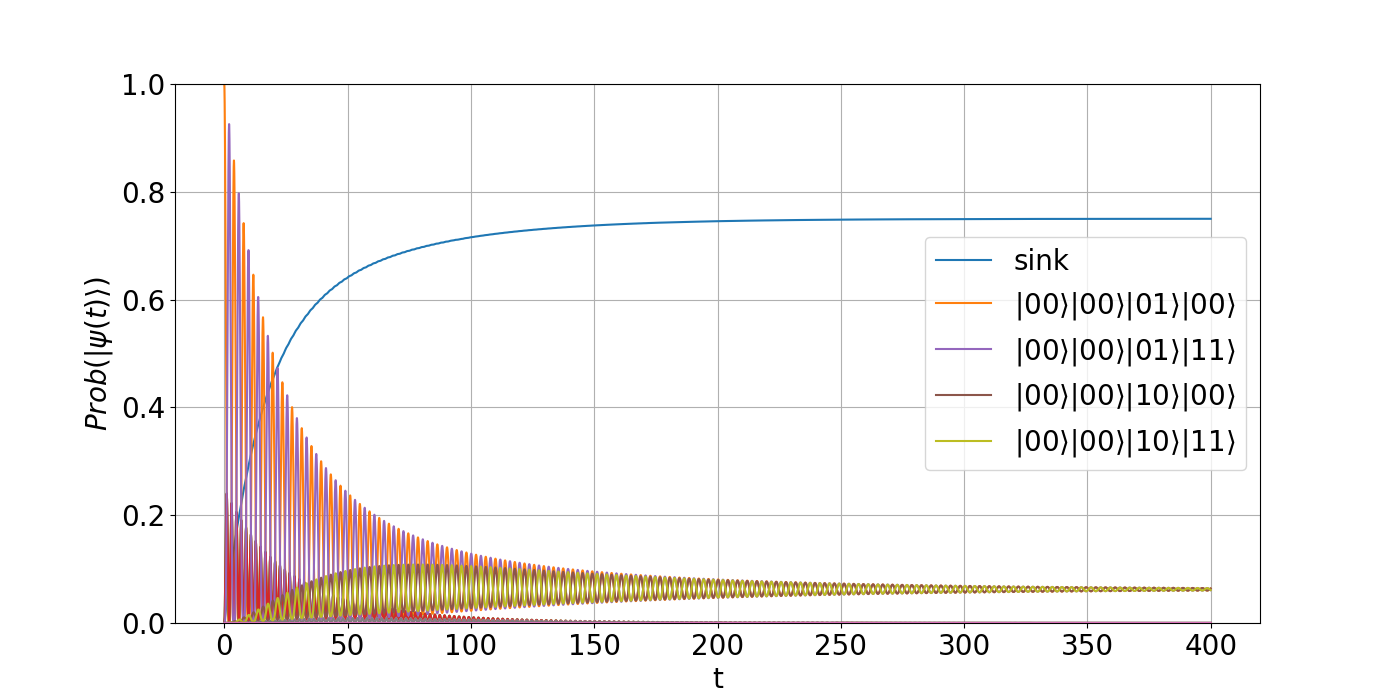}
			Calculation using state vectors.
			\label{fig:test_TCH_psi_1_40000_leak.png} 
		\end{minipage}
		\caption{The case of two two-level atoms in two cavities. $dt=0,01, \hbar=1, g_1=g_2=g=2,\mu=0.8,r=0.4, \omega = 1$ and initial state $|\Psi(0)\rangle=|00\rangle|00\rangle|01\rangle|00\rangle$ -- there are no photons in the cavity, and one of the atoms is excited. The total number of possible states is 24.}
		\label{fig:TCH_2_2}
	\end{figure} 
	
	In this model, we obtained a selection of the dark state for atoms that allow spatial motions (the so-called {\it black} states):
	\begin{equation}
		|Black\rangle = \frac{1}{4}|00\rangle|00\rangle\left(|01\rangle|00\rangle -|10\rangle|00\rangle + |01\rangle|11\rangle - |10\rangle|11\rangle\right)
		\label{equation5.8}
	\end{equation}
	
	Our algorithm is consistent with the results calculated using density matrices. For systems with more than 1000 ground states, we cannot use the density matrix to compute finite time results. But using our algorithm can help us find dark or black states of higher dimensionality.
	
	At the same time, as the number of atoms in the system increases, finding all possible states and filling in the Hamiltonian also presents a difficulty. Because the search for all possible states into which a transition from the initial state is possible requires a constant search and modification of existing data. We used a hash table and dictionary structure to generate the required data in a short time.
	
	For 4 atoms we got the following result
	\begin{figure}[H]
		\centering
		\begin{minipage}[t]{0.49\linewidth}	
			\centering
			\includegraphics[width=3.3in]{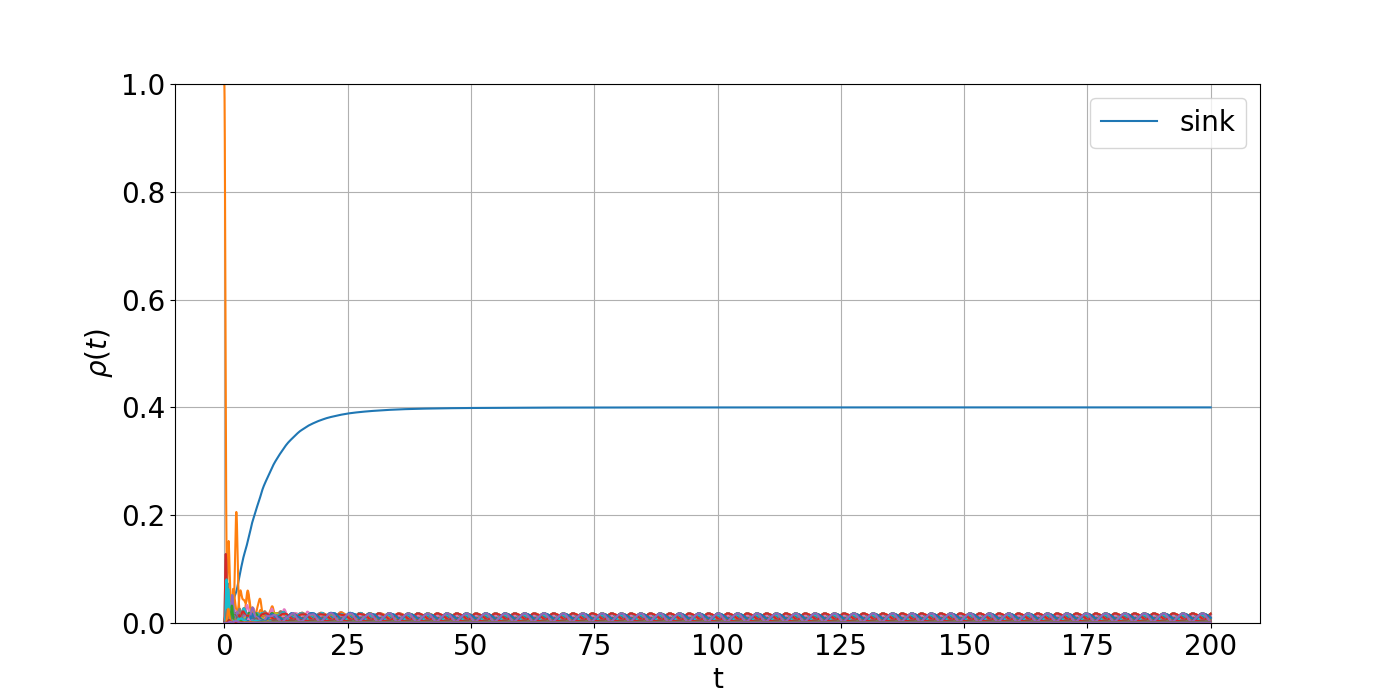}
			Calculation using density matrix. $\gamma=0.5$.
			\label{fig:image_TCH_rho_20000_000000110000_leak.png} 
		\end{minipage}\begin{minipage}[t]{0.49\linewidth}
			\centering
			\includegraphics[width=3.3in]{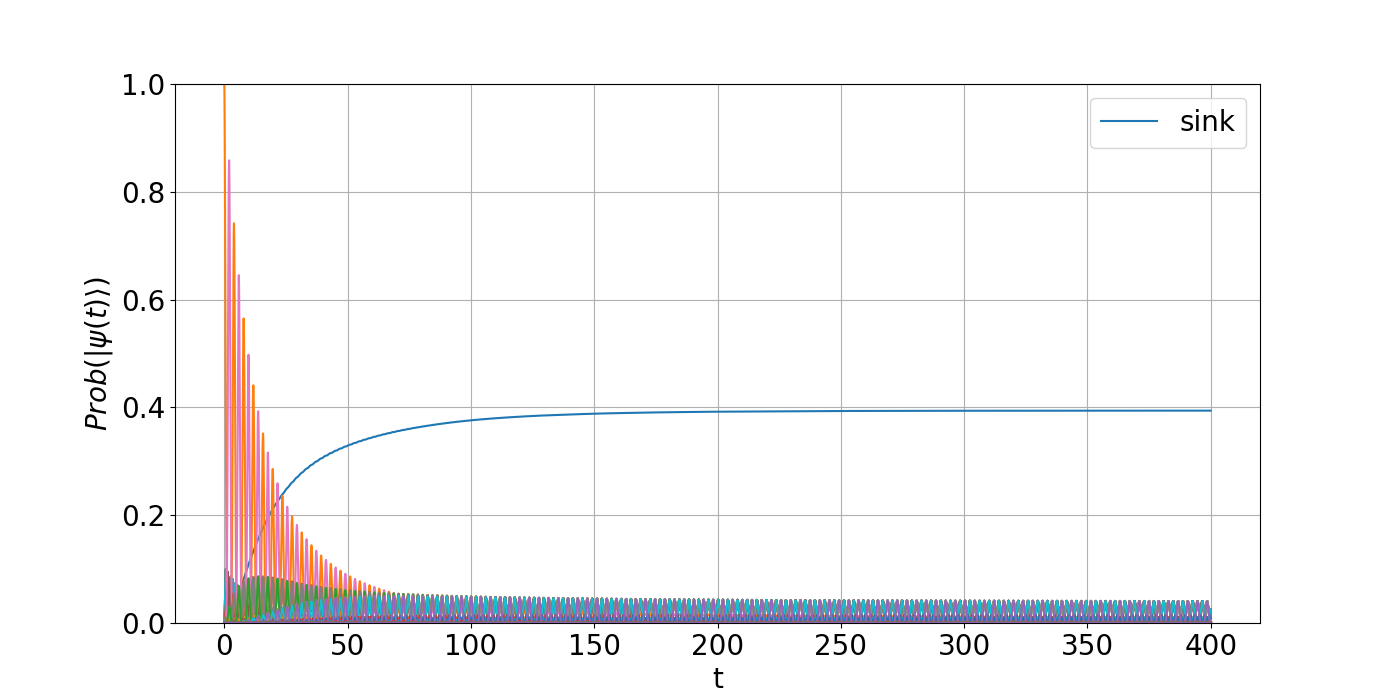}
			Calculation using state vectors.
			\label{fig:image_TCH_40000_000000110000_leak.png} 
		\end{minipage}
		\caption{Four two-level atoms in two cavities. $dt=0,01, \hbar=1, g_1=g_2=g=2,\mu=0.8,r=0.4, \omega = 1$ and initial state $|\Psi(0)\rangle=|00\rangle|00\rangle|0011\rangle|0000\rangle$ -- there are no photons in the cavity, and two of the atoms is excited. The total number of possible states is 512.}
		\label{fig:TCH_2_4}
	\end{figure} 
	
	For 6 atoms we got the following result
	\begin{figure}[H]
		\centering
		\centering
		\includegraphics[width=3.3in]{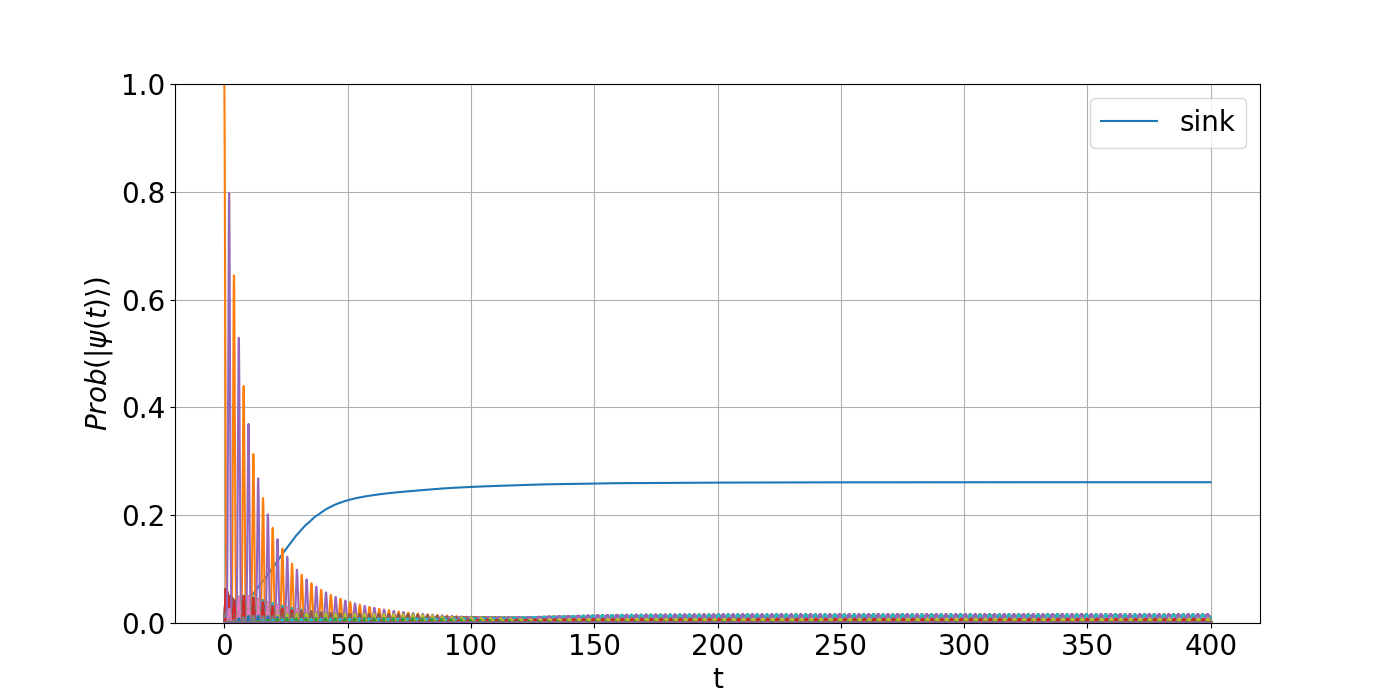}
		\caption{6 two-level atoms in two cavities. $dt=0,01, \hbar=1, g_1=g_2=g=2,\mu=0.8,r=0.4, \omega = 1$ and initial state $|\Psi(0)\rangle=|00\rangle|00\rangle|000111\rangle|000000\rangle$ -- there are no photons in the cavity, and three of the atoms is excited. The total number of possible states is 10240.}
		\label{fig:TCH_2_6}
	\end{figure} 
	In all cases, our algorithm singled out the dark component of the initial state in exactly the same way as the QME.
	To compare the complexity of modeling according to our algorithm and according to the QME, we received the following comparative table
	
	\noindent
	\begin{table}[H]
		\centering
		\begin{tabular}{ccccccc}
			\toprule
			$N$ & $Initial\ state$ & $Dim$ & $MC_\rho$ & $time_\rho$ & $MC_v$ & $time_v$\\
			\midrule
			2 & $|01\rangle$ & 24 & 4.5 KB  & <0.00001 s & 192 B & <0.00001 s \\
			4 & $|0011\rangle$ & 512  & 2 MB & 0.1231 s & 4 KB & 0.00138 s \\
			6 & $|000111\rangle$ & 10240 & 0.78 GB & 900.508 s & 80 KB  & 0.04007 s \\
			8 & $|00001111\rangle$ & 196864 & 288.75 GB & --- & 1.5 MB & 1.36459 s \\
			10 & $|0000011111\rangle$ & 3684352 & 98.76 TB & --- & 28.11 MB & $>$10 s \\
			\bottomrule
		\end{tabular}
		\caption{Comparison table of two algorithms.}
		\label{tab:table_of_the_two_algorithms}
	\end{table}
	Here $N$ is the number of atoms, $Initial state$ is the initial state of the atom, where half of the atoms are excited, $Dim$ is the number of possible states, $MC_\rho$ is the minimum memory required to use the density matrix, $MC_v$ is the memory required to use the state vector calculation, $time_\rho$ is the time required to iterate once using the density matrix, $time_v$ is the time required to iterate once using the state vector. The Hamiltonian is stored as a sparse matrix, which is less than the memory required to use a dense matrix.
	
	The temporary results of one iteration in the table \ref{tab:table_of_the_two_algorithms} are 100 iterations in total in the program, and the average value is taken from 5 experiments. In the case of 10 atoms, it takes about half an hour to find all possible states and fill in the Hamiltonian. The number of possible states has reached a staggering 3 million. In the case of 12 atoms, even if a hash table and dictionary structure was used for the lookup, it is difficult to complete it in a reasonable amount of time.
	
	We see that it is impossible to use the density matrix in the case of 8 atoms. But it can still be calculated using the state vector. With the help of parallel tools such as MPI, the efficiency of the program will increase, but for 10 atoms, the calculation time of one iteration exceeds 10 seconds. In the case of $dt=0.01$, we generally need to repeat thousands or even tens of thousands of times. This is no longer acceptable for computing on a single computer.
	
	Thus, the limit of the density matrix evolution method can be increased from 4 atoms to 8-10 using the state vector evolution method. In the case of 12 atoms, the dimension can reach tens of millions. Even with the help of a supercomputer, this cannot be done in a reasonable amount of time. This limitation means that for more than 10 atoms, completely different methods must be used to work with the pure state vector.
	
	\section{Conclusions}
	
	We have constructed an efficient replacement of the quantum basic equation in the form of an iterative algorithm scheme that specifies the time variation of the pure state vector $|\Psi(t)\rangle$. This scheme covers both the case of QED in cavities, possibly not ideal, and the case of a distributed atomic system in open space.
	
	Our algorithmic scheme has two main advantages over the quantum master equation. The first is that this scheme fully takes into account the possible memory of the environment in the form of the return of an emitted photon or the exchange of photons between different locations. The second advantage is that our scheme operates with a pure state, and not with a density matrix, like the QME. For complex systems consisting of hundreds of atoms, this advantage is very important, since the computer memory required to store the matrix grows as the square of the length of the vector.
	
	The proposed scheme is well suited for parallelization and implementation on distributed computing systems. Finally, it is methodologically simpler: the probability here is not divided into classical and quantum, as when working with density matrices. This algorithmic scheme is not as universal as the QME, since it is applicable only to the case of the exchange of photons or phonons with the environment, and does not cover other types of decoherence, for example, atomic operators of the $\sigma^+\sigma$ type, which occur in some relaxation models . However, the factor of exchange of bosons with the environment is quite important for complex non-Markovian processes with a meaningful scenario, and here our scheme has a good perspective.

\end{document}